\documentclass[12pt]{article}
\newcommand{\be}[1]{\begin{equation}\label{#1}}
    \newcommand{\ba}[1]{\begin{eqnarray}\label{#1}}

    \newcommand{\pa}[1]{\left(#1\right)}
    \newcommand{\paq}[1]{\left[#1\right]}

    \def\ee{\end{equation}}
    \def\ea{\end{eqnarray}}

\usepackage{graphicx,latexsym}
\usepackage{graphicx,epsf}
\usepackage{color}
\usepackage{epsfig}
\usepackage{amsmath}
\usepackage[T1]{fontenc}
\usepackage[utf8]{inputenc}
\usepackage{authblk}
\begin{document}
\title{Vacuum energy, Standard Model physics  	
and the $750\; \rm{GeV}$ Diphoton Excess at the LHC }
\author[1,2]{Alexander Y. Kamenshchik\thanks{Alexander.Kamenshchik@bo.infn.it}}
\author[2,3]{Alexei A. Starobinsky\thanks{alstar@landau.ac.ru}}
\author[1]{Alessandro Tronconi\thanks{Alessandro.Tronconi@bo.infn.it}}
\author[1]{Gian Paolo Vacca\thanks{Vacca@bo.infn.it}}
\author[1]{Giovanni Venturi\thanks{Giovanni.Venturi@bo.infn.it}}
\affil[1]{Dipartimento di Fisica e Astronomia and INFN, Via Irnerio 46,40126 Bologna,
Italy}
\affil[2]{L.D. Landau Institute for Theoretical Physics of the Russian
Academy of Sciences, Kosygin str. 2, 119334 Moscow, Russia}
\affil[3]{Bogoliubov Laboratory for Theoretical Physics, Joint Institute for Nuclear Research, Dubna, Moscow Region, 141980, Russia}
\date{}

\maketitle

\begin{abstract}
The conditions for the cancellation of one loop contributions to
vacuum energy (both U.V. divergent and finite) coming from the
known (Standard Model) fields and the field associated with the hypothetical 
750 Gev diphoton excess are examined. 
Depending on the nature of this latter resonance we find that in order to satisfy 
the constraints (sum rules) different additional,   minimal field contributions are required, leading to yet to be observed particles.
\end{abstract}

\section{Introduction}
%%%%%%%%%%%%%%%%%%%%%%%%%%%%%
Almost seventy years ago Pauli~\cite{Pauli} suggested that the
vacuum (zero-point) energies of all existing fermions and bosons
compensate each other. This possibility is based on the fact that
the vacuum energy of fermions has a negative sign whereas that of
bosons has a positive one. We note that such an idea is realised in
a highly constrained way in supersymmetric models, although
supersymmetry breaking must be present at probed energies in order
to explain the observed data. Subsequently in a series of papers
Zeldovich~\cite{zeld} connected the vacuum energy to the
cosmological constant, however rather than eliminating the
divergences through a boson-fermion cancellation, he suggested a
Pauli-Villars regularisation of all divergences by introducing a
number  of massive regulator fields. Covariant regularisation of
all contributions then leads to finite values for both the energy
density $\varepsilon$ and (negative) pressure $p$ corresponding to a cosmological
constant, i.e. connected by the equation of state $p =-\varepsilon$. 

The possibility that such a finite (renormalized) vacuum energy acts as a source for the gravitational dynamics
leads to the cosmological constant puzzle of modern physics. 

Developing an approach based on the
combination of the above two ideas, that is, as Pauli suggested the {\it existing}
boson and fermion states (fields) should provide the exact cancellation of all the
ultraviolet divergences in the vacuum energy while, as Zeldovich
suggested, the remaining finite part of the vacuum energy should lead to an
effective cosmological constant, is a non trivial task.

Indeed the original Pauli idea should be extended in a general QFT context where most generally
the full structure of all contributions would require solving the dynamics for all interactions. 

One possible approach to the problem would be to consider a fundamental QFT setup defined below a UV scale lower than the Planck mass
(omitting for the time being quantum gravity effects).
This requires choosing a bare action from which to compute the full effective action including all eventual power-like divergences~\footnote{
In a free theory context a way to see how and under which assumptions one can derive in a Wilsonian framework the Zeldovich formula in Eq.~(\ref{en}) was discussed in~\cite{VZ} in Sect. $3.2$.} which is a complicated task.

In order to make the problem more tractable, it is better to consider
a traditional perturbative approach, that is a loop expansion computed from the effective action.
if we then consider an expansion in powers of $\hbar$, different loop orders can therefore contribute different powers of $\hbar$.
In particular the contributions to the vacuum energy (field independent part of the effective action)
to the lowest order in $\hbar$ are obtained from the one loop contributions related to the massive fields employing their ``physical masses". 

In our specific approximation we shall require that the cancellations, in order to obtain a finite, actually negligible, cosmological constant, 
should be imposed order by order in $\hbar$ and here we shall limit ourselves to an analysis of the $O(\hbar)$ (lowest order) sum rules. 
This is in agreement with the fact that in a semiclassical analysis, $\hbar$ being small, gives the dominant corrections.
In this framework we shall ignore, as a further approximation, contributions from bound states.

In a previous work we have already examined~\cite{we} the problem of the cancellation at order $O(\hbar)$
of the UV divergences of the vacuum energy for both Minkowski and de
Sitter space-times and formulated the conditions (sum rules) for the
cancellation of all divergences. 
These conditions led to strong restrictions on the spectra of possible fields. In paper \cite{we1} we applied such considerations to the observed
particles of the Standard Model (SM) and also  studied the finite part of the vacuum
energy. In particular, also the possibility of a cancellation for this last contribution, 
so as to obtain a result compatible with the observed value of the cosmological constant
(almost zero with respect to SM particle masses). 
We showed \cite{we1} that it was impossible to construct a minimal extension of the
SM by finding a set of boson fields which, besides cancelling the ultraviolet
divergencies, could compensate the residual huge contribution of the known
fermion and boson fields of the Standard Model to the finite part of the
vacuum energy density.

On the other hand we  found that the addition of at least one massive fermion field
was sufficient for the existence of a suitable set of boson fields which would permit
such cancellations and obtained their allowed mass intervals. 
This result was by itself very suggestive since in extensions of the SM often new
extra fermions are considered, independently of any cancellation requirement.
Further, on examining
one of the simplest SM extensions satisfying the constraints we 
found that the mass range of the lightest massive boson was
compatible with the Higgs mass bounds, which were known at the moment of the publication of the paper 
\cite{we1}. 

Subsequently some very  important results were obtained in  experimental particle physics. First of all the Higgs boson was discovered \cite{Higgs}. This long awaited event  accomplished the experimental confirmation of the so called Standard Model. Quite recently another event has excited  the physics community: the observed diphoton excess at 750 \rm{GeV} \cite{diphoton}. 
This excess, if confirmed, could be interpreted as an indication for the existence of a new heavy elementary or composite particle with  a mass of the order of 750 \rm{GeV}. Not much is known about such a possible state, including if it may have a narrow or large width, in which case other decay channels beyond the SM photon and gluons (typically leading to small width contributions at loop level) could be invoked, eventually related to new particles or composite states, related to dark matter.
This has led to  a rather frenetic theoretical activity \cite{theor}.

In the present paper 
we wish to apply the methods and ideas, developed in the preceding papers \cite{we,we1} to the analysis of this new experimental situation. 

The structure of the paper is the following: in the second section we write down the conditions for the cancellations of divergent and finite contributions  between fermion and boson fields to the vacuum energy, we  briefly present the results of our preceding paper \cite{we1} and generalise some formulae by adding more constraints to the model; in the third section we shall take into account different possibilities for the new hypothetical $750\;{\rm GeV}$ particle and the resulting possible structures of the spectrum for elementary particles beyond the Standard Model. The last section contains some concluding remarks. 

%%%%%%%%%%%%%%%%%%%%%%%%%%%%%%%%%%%%%%%
\section{Vacuum energy and the balance between the fermion and boson fields}

As previously discussed we start by considering the contribution of order $O(\hbar)$ to the vacuum energy of the propagating massive particles of a fundamental theory.
One knows that the vacuum energy of the harmonic oscillator is equal to $\frac{\hbar\omega}{2}$. If one has  a massive  field with  mass $m$, then $\omega =\sqrt{k^2c^2 + m^2c^4}$, where $k$ is the wave number.  In the following we shall set $\hbar=1$ and $c=1$. The energy density of the vacuum energy of a scalar field, treated as oscillators with all possible momenta is given by the divergent integral \cite{zeld}:
\begin{equation}
\varepsilon=\frac12\int d^3k\sqrt{k^2+m^2} = 2\pi\int_0^{\infty}dk k^2 \sqrt{k^2+m^2}. 
\label{en}
 \end{equation}
We can regularise this integral by introducing a cutoff $\Lambda$. In this case
\begin{eqnarray}
&&\varepsilon=2\pi\int_0^{\Lambda}dk k^2 \sqrt{k^2+m^2}\nonumber \\
&&=2\pi m^4\left[\frac{\Lambda}{8m}\left(\frac{2\Lambda^2+1}{m^2}\right)\sqrt{\frac{\Lambda^2}{m^2}+1}-\frac18\ln\left(\frac{\Lambda}{m}+\sqrt{\frac{\Lambda^2}{m^2}+1}\right)\right].
\label{en1} 
 \end{eqnarray}
On expanding this expression with respect to the small parameter $\frac{m}{\Lambda}$, one obtains
\be{en2}
\varepsilon=\frac{\pi}{2}\Lambda^4+\frac{\pi}{2}\Lambda^2m^2+\frac{\pi}{16}m^4(1-4\ln2)-\frac{\pi}{4}m^4\ln\frac{\Lambda}{m}+o\left(\frac{m}{\Lambda}\right).
\ee
The contribution of one fermion degree of freedom coincides with that of Eq. (\ref{en})  with the opposite sign. 
It now follows from Eq. (\ref{en2}) that to cancel the quartic ultraviolet divergences, proportional to $\Lambda^4$, one has to have equal numbers of boson and fermion degrees of freedom:
\begin{equation}
N_B=N_F. 
\label{en0}
\end{equation}
The conditions for the cancellation of quadratic and
logarithmic divergences  are
\begin{equation}
\sum m_S^2 +3 \sum m_V^2 = 2 \sum m_F^2
\label{quad}
\end{equation}
and
\begin{equation}
\sum m_S^4 +3 \sum m_V^4 = 2 \sum m_F^4\,,
\label{log}
\end{equation}
respectively.
Here the subscripts $S$, $V$ and $F$ denote scalar, massive vector and massive
spinor Majorana fields respectively (for Dirac fields it is sufficient to put a $4$
instead of $2$ on the right-hand sides of Eqs. (\ref{quad}) and (\ref{log})).
For the case such that the conditions (\ref{en0}), (\ref{quad}) and (\ref{log}) are satisfied the remaining finite part of the vacuum energy density is  
equal to
\begin{equation}
\varepsilon_{\rm finite}= \sum m_S^4\ln m_s +3 \sum m_V^4 \ln m_V - 2 \sum m_F^4\ln m_F.
\label{en-finite}
\end{equation}
Let us now calculate the vacuum pressure, this pressure is given by the formula \cite{zeld}
\begin{equation}
p=\frac{2\pi}{3}\int_0^{\infty}dk\frac{k^4}{\sqrt{k^2+m^2}}.
\label{pres-vac}
\end{equation}
On introducing the cutoff $\Lambda$ we have
\begin{eqnarray}
&&p=\frac{2\pi}{3}\int_0^{\Lambda}dk\frac{k^4}{\sqrt{k^2+m^2}}\nonumber \\
&&=\frac{2\pi}{3}m^4\left[\frac18\frac{\Lambda}{m}\left(\frac{2\Lambda^2}{m^2}\right)\sqrt{\frac{\Lambda^2}{m^2}+1}-\frac{\Lambda}{m}\sqrt{\frac{\Lambda^2}{m^2}+1}\right.\nonumber \\
&&\left.+\frac38\ln\left(\frac{\Lambda}{m}+\sqrt{\frac{\Lambda^2}{m^2}+1}\right)\right].
\label{pres-vac1}
\end{eqnarray}
On expanding this expression with respect to the small parameter $\frac{m}{\Lambda}$, we obtain
\be{pres-vac2}
p=\frac{\pi}{6}\Lambda^4-\frac{\pi}{6}\Lambda^2m^2-\frac{7\pi}{48}m^4+\frac{\pi}{4}\ln 2+\frac{\pi}{4}m^4\ln\frac{\Lambda}{m}+o\left(\frac{m}{\Lambda}\right).
\ee
Then, on comparing the expressions (\ref{en2}) and (\ref{pres-vac2}), we see that the quartic divergence satisfies the  equation of state  
for  radiation $p=\frac13\varepsilon$, the quadratic divergence satisfies the equation of state $p=-\frac13\varepsilon$, which sometimes is identified with the so called string gas (see e.g. \cite{string}), while the logarithmic divergence behaves as a cosmological constant with $p=-\varepsilon$. If all these divergences  cancel, then the finite part of the pressure is 
\begin{equation}
p_{\rm finite}=-(\sum m_s^4\ln m_s +3 \sum m_V^4 \ln m_V - 2 \sum m_F^4\ln m_F)
\label{p-fin}
\end{equation}
which also behaves like a cosmological constant.

The requirement that the
finite part of the vacuum energy, i.e. of the observable effective cosmological constant,  is very small
compared with SM masses suggests that we also need a compensation
between the finite parts of fermion and boson vacuum energies,
thus obtaining
\begin{equation}
\sum m_S^4\ln m_S +3 \sum m_V^4 \ln m_V - 2 \sum m_F^4\ln m_F =0.
\label{en-finite1}
\end{equation}

As is known the observed number of fermion degrees of freedom in the Standard
Model is much bigger than the number of boson degrees of freedom \cite{particles}.
Indeed $N_F$ is equal to $96$ (if we consider the neutrinos as massive particles)
while the number of boson degrees of freedom, carried by the photon,
the gluons, $W^{\pm}$ and $Z^0$ bosons and by the Higgs boson  is equal to $28$.
Thus we need an additional 68 boson degrees of freedom. 
At this point it is natural to search for some minimal extension of the Standard Model, which
does not modify the fermion degrees of freedom while just adding some hypothetical bosons.

The main  result of paper \cite{we1} was a proof that, within the given framework, such an
extension did not exist. In other words, we showed that on
introducing new boson fields, which provide the cancellation of
the ultraviolet divergences in the vacuum energy density, the finite
part of the effective cosmological constant is always positive and
of order of the mass of the top quark to the fourth power, which is
much bigger than the value of the effective cosmological constant
compatible with  cosmological observations. This led to the
necessity of introducing new heavy fermions. Indeed we  found
explicit realisations with zero finite energy by introducing at
least one fermion with a suitable mass.

Let us now  sketch briefly the approach of the preceding paper \cite{we1} 
which will also be used  in the present paper.

After a general analysis,  for the sake of simplicity, we considered an
explicit minimal extension of the SM with a few massive bosons and weakly coupled,
practically massless others so as to satisfy the requirement $N_B=N_F$.
Such a possibility is viable in effective action approaches and, for example, has been
considered recently in scenarios such as unparticle physics~\cite{georgi}.
In this minimal framework we  analysed the boson masses allowed by the
cancellation constraints.

Let us suppose that the contributions of fermions (or bosons) to the constraint equations  
(\ref{quad}), (\ref{log}) and (\ref{en-finite1}) are bigger than those of bosons (or fermions).
In the Standard Model before the observation of the 750 \rm{GeV} Diphoton Excess at the LHC
the fermions dominated. However, with this last observation, the situation in inverted: it is now bosons that dominate. For definiteness lets  talk about the dominance of the fermions, although it is not essential for the formalism. Thus, we take  the differences between the contributions of fermions and bosons into the constraint equations 
 to be  positive and  call them $R^2$, $h$ and $L$ respectively. Then, if we call $x_i$ the masses squared of the boson degrees of freedom which we wish to add to create the balance, we can write down the constraint equations as 

\begin{equation}
\sum_{i=1}^{n} x_i^2 = R^2, \label{constraint1}
\end{equation}
\begin{equation}
\sum_{i=1}^{n} x_i = h, \label{constraint2}
\end{equation}
\begin{equation}
\phi\equiv\sum_{i=1}^{n} x_i^2 \ln x_i = L, \label{constraint3}
\end{equation}
where $n$ is the number of the boson degrees of freedom. 

The  constraints  (\ref{constraint1}), (\ref{constraint2})  have a simple geometrical sense~\cite{we}: they describe a sphere and a plane in an $n$-dimensional space and their intersection {\cal S} is an $(n-2)$--dimensional sphere, eventually to be sliced on the positivity boundaries of the $x_i$. The distance of the plane from the origin of the coordinates is $h/\sqrt{n}$. In order to have an intersection between the sphere of radius $R$ and the plane it is
then necessary to have
\begin{equation}
n > \frac{h^2}{R^2} 
\label{condition5}
\end{equation}
In general it is convenient to introduce the integer value $n_0$ for such a threshold
\begin{equation}
n_0 = \left[\frac{h^2}{R^2} + 1 \right],
\label{n0}
\end{equation}
so that $n\ge n_0$ is the requirement to have a non empty {\cal S} (here the square brackets  denote the integer part of a number). Further, the sphere $S$ should contain  points such that all the coordinates $x_i$ are positive. This condition is given by the inequality 
\begin{equation}
h^2>R^2
\label{posit}
\end{equation}
and geometrically is equivalent to the requirement that the hyperplane (\ref{constraint2}) intersect the axis of the n-dimensional space outside the hypersphere (\ref{constraint1}). 

The conditions (\ref{condition5}-\ref{posit}) for the existence of physical solutions to the geometrical constraint equations (\ref{constraint1}-\ref{constraint2}) can be reduced to the following inequality
\be{cons1+2}
R^2<h^2<nR^2.
\ee 
Then, in principle, if (\ref{posit}) is verified, one can find a multiplet of massive bosonic particles satisfying the geometrical constraints.\\
In \cite{we} we calculated the maximum and minimum values for the  $x_i$'s satisfying the geometrical equations. When the constraint (\ref{constraint3}) is added such values also give an estimate of the interval on which $x_i$'s vary.
When  more general model building is considered additional hypothesis are added to the above constraints (\ref{constraint1}-\ref{constraint3}). For example one may ask for cancellations among particle multiplets with different multiplicities $w_i$'s. In such a case the above formulae should be modified accordingly. In particular the geometrical equations become
\be{geomulti}
\sum_{i=1}^{n} w_i x_i^2 = R^2, \quad \sum_{i=1}^{n} w_i x_i = h
\ee
where $n$ is the number of particles in the multiplet and, by redefining $\sqrt{w_i}x_1\equiv y_i$ one is led to slightly modified geometrical constraints with the same hypersphere but a different hyperplane. The distance between the plane and the origin $y_i=0$ is now $d=h/\sqrt{\sum_{i=1}^{n}w_i}$ and the condition (\ref{condition5}) becomes
\be{condition5b}
\tilde n>\frac{h^2}{R^2}
\ee
where $\tilde n$ is the number of degrees of freedom. In order to have positive solutions from the intersection between the sphere and the plane we now have 
\be{positb}
h^2>w_{\rm min}R^2
\ee
where $w_{\rm min}$ is the minimum multiplicity we consider in the multiplet. Then, if the condition
\be{cons1+2b}
w_{\rm min}R^2<h^2<\tilde n R^2
\ee
is satisfied we have physically acceptable solutions from the intersection (\ref{geomulti}).\\
One can calculate, for each $\bar x$ solution of (\ref{geomulti}), how their maximum and minimum ($\bar x_M$ and $\bar x_m$ respectively) vary as a function of the corresponding multiplicity $\bar w$. We find:
\be{minmax}
\bar x_{M,m}=\frac{h}{\tilde n}\pm \sqrt{\frac{\tilde nR^2-h^2}{\tilde n^2}}\sqrt{\frac{\tilde n}{\bar w}-1}
\ee
and we observe that, correspondingly, the other masses $x_i$ in the multiplet are equal.\\
For a given value of $h,R$ and a suitable choice of $\tilde n$ we find that a multiplicity larger than one leads to a smaller maximum and a bigger minimum. Let us note that the minimum is positive if 
\be{posmin}
n<\frac{h^2}{R^2}+w\Rightarrow n\le \paq{\frac{h^2}{R^2}}+w=n_0+w-1,
\ee
and then in order to find very massive particles in the intersection between the sphere and the plane one must restrict the analysis to the case $n=n_0$.\\
The problem of finding the extrema of the $x_i$'s can be generalised and the whole set of three constraints (\ref{constraint1}-\ref{constraint3}) taken into account. For such a case, let us suppose we need the extrema for $x_1$, then one can introduce the auxiliary function 
\be{aux1}
f(\{x_i\})\equiv x_1+\alpha\pa{\sum_{i=1}^{n} w_1 x_i^2 \ln x_i-l} - \lambda \left(\sum_{i=1}^{n} w_i x_i^2 - R^2\right) - \mu
\left(\sum_{i=1}^{n} w_i x_i - h\right)
\ee
where $\alpha$, $\lambda$ and $\mu$ are three Lagrange multipliers. The condition $\partial f/\partial x_i=0$ takes the form
\be{derlag}
\frac{\delta_{1i}}{w_1}+\alpha\pa{2x_i \ln x_i+x_i}+2\lambda x_i+\mu=0
\ee
and is independent on $w_i$ for $i\neq 1$. One then observes that the max/min values for $x_1$ are in the multiplets formed by 3 different masses: $\bar x_1$, $\bar x_2$, $\bar x_3$ satisfying the constraints (\ref{constraint1}-\ref{constraint3}):
\be{consminmax}\left\{
\begin{array}{l}
w_1 \bar x_1+\tilde n_2 \bar x_2+\pa{\tilde n-w1-\tilde n_2}\bar x_3=h\\
\\
w_1 \bar x_1^2+\tilde n_2 \bar x_2^2+\pa{\tilde n-w1-\tilde n_2}\bar x_3^2=R^2\\
\\
w_1 \bar x_1^2\ln \bar x_1+\tilde n_2 \bar x_2^2\ln \bar x_2+\pa{\tilde n-w1-\tilde n_2}\bar x_3^2\ln \bar x_3=L
\end{array}\right.
\ee
where $w_1$ is the multiplicity associated with $x_1$, $\tilde n$ is the total number of d.o.f. chosen to satisfy the constraints and $\tilde n_2$ is number of d.o.f. with mass $\bar x_2$. Depending on the multiplicities $w_i$, $\tilde n_2$ may take a few possible values between $1$ and $\tilde n-w_1-1$. The solutions of the system (\ref{consminmax}) can be found numerically on varying $\tilde n_2$ and then the max/min values for $x_1$ can be determined. Let us note that the above approach can be extended to the case of hybrid particle multiplets containing both boson and fermion d.o.f. on switching the relative sign of the $w_i$'s. In this latter case  eq. (\ref{derlag}) for $i\neq 1$ is unchanged and one can still conclude that the multiplet consists of  3 masses.

%%%%%%%%%%%
In order to now see when the condition (\ref{constraint3}) can also be 
satisfied, it is convenient to calculate the minimum value  of the
function $\phi=\frac{1}{2}\sum_{i=1}^{n} x_i^2 \ln x_i^2$ on the constraint
surface {\cal S}.
 Let us consider an auxiliary function
\be{Lagrange}
F(\{x_i\}) =  \frac{1}{2}\sum_{i=1}^{n} x_i^2 \ln x_i^2 - \lambda \left(\sum_{i=1}^{n} x_i^2 - R^2\right) - \mu
\left(\sum_{i=1}^{n} x_i - h\right),\ee
where $\lambda$ and $\mu$ are the Lagrange multipliers.
The search for the extrema of the function $F$ implies we should equate its
derivatives with respect to $x_i, \lambda$ and $\mu$ to zero. These
last two conditions $\partial F/\partial \lambda$ and $\partial F/
\partial \mu$ again give the  constraints  (\ref{constraint1}), (\ref{constraint2}).
Differentiation with respect to $x_i$ gives the system of equations:
\begin{equation}
x_i \ln x_i^2 +x_i - 2\lambda x_i - \mu = 0,\ \
i=1,\cdots,n .\label{system}
\end{equation}
Without loss of generality we can choose $x_1 \neq x_2$. Indeed it is
possible to have $x_1 = \cdots = x_n$ if and only if $h^2/n^2 =
R^2$, but this is a degenerate case, when the sphere and the plane
touch each other in only one point.

On substituting the values of $x_1$ and $x_2$ into the first two
equations of the system (\ref{system}), one obtains
$\bar{\lambda}$ and $\bar{\mu}$ as functions of $x_1$ and
$x_2$:
\begin{equation}
\bar{\lambda} = \frac{1}{2} + \frac{x_1\ln x_1^2 - x_2\ln x_2^2}{2\pa{x_1 - x_2}},\quad
\bar{\mu} = \frac{x_1 x_2 (\ln x_2^2 - \ln x_1^2)}{x_1 - x_2}.
\label{lambdamu}
\end{equation}

Let us now suppose that
$\bar{x}_1,\cdots,\bar{x}_n,\bar{\lambda},\bar{\mu}$ are a solution
of the system (\ref{system}) on {\cal S}, i.e. with the  constraints
(\ref{constraint1}) and (\ref{constraint2})  already satisfied. On
then substituting these values of $\bar{\lambda}$ and $\bar{\mu}$
into the $n-2$ remaining equations of the system (\ref{system}) one
can easily see that a solution is given by $x_1=x_3 = x_4 = \cdots =
x_{k+1}$ and $x_2=x_n=x_{n-1}=\cdots=x_{k+2}$. This solution is a
stationary point of the function $F$, or in other words the
conditional stationary point of the function $\phi$. Such a
solution, with $k$ coordinates having the value $\bar{x}_1=x$ and
the remaining $n-k$ coordinates the other value $\bar{x}_2=y$, is
given, as function of $k$ and $n$, by
\begin{equation}
x=x(k,n) = \frac{h}{n} + \sqrt{\frac{R^2(n-k)}{nk} - \frac{h^2(n-k)}{n^2k}}\,,
\label{solx1}
\end{equation}
\begin{equation}
y=y(k,n) = \frac{h}{n} - \sqrt{\frac{R^2 k}{n(n-k)} - \frac{h^2 k}{n^2(n-k)}}\,,
\label{soly1}
\end{equation}
where $1 \leq k \leq n-1$.

The values of $x$ given by Eq. (\ref{solx1}) are always positive,
while the values of $y$ can be negative.
It is easy to show that the condition for the positivity of $y$ is
\begin{equation}
k < n_0 \leq n \,.
\label{positive}
\end{equation}
We have seen that points of the type described above always
satisfy the stationarity conditions  (\ref{system}) on the constraint surface.
This does not mean that stationary points of other types cannot exist.
Indeed, the  analysis of the structure of Eq. (\ref{system}) shows
that, in principle,
stationary points whose coordinates $x_i$ have {\it three} different
values can exist. However, if such points exist, at least one of these three
values is negative and, hence, is of no interest to us. Thus, the minimum of
the function $\phi$ can be reached {\it only} for the stationary points having
the coordinates (\ref{solx1}), (\ref{soly1}) or on the boundary of the
positivity region, where at least one of the coordinates $x_i$ is equal to
zero. For this last case the problem is reduced to one with a lower dimensionality than $n$.

If $n=n_0$ (the smallest possible value for the dimensionality
of $n$) we notice that on the surface {\cal S} all the $x_i$ have positive values.
Thus, the maximum and minimum values of the function $\phi$ on the constraint
surface are obtained only for one of the pairs of points with the coordinates $x$
and $y$ (see formulae (\ref{solx1}), (\ref{soly1})).

Furthermore the following more general statement is true: for a given $n \ge n_0$
the maximum value of the function $\phi$ corresponds to $k = 1$
while its minimum value corresponds to $k = n-1 $.
To prove this one may compute the derivatives of the function
\begin{equation}
\phi_1(k,n) = \frac{k\, x^2}{2}\ln x^2 + \frac{(n-k)\, y^2}{2} \ln y^2,
\label{phi1}
\end{equation}
with respect to $k$ and n. It can be shown that $d\phi_1/dk<0$
and $d\phi_1/dn>0$ for the range of possible physical values of $k$ and $n$.
In particular this means that the function $\phi_1(k,n)$ decreases with
increasing $k$ and has its minimum value at $k = n_0-1$ and $n=n_0$. Once one chooses $n$ such that it satisfies the geometrical constraints then the minimum and the maximum values for $\phi_1(k,n)$ are respectively
\be{minim}
\phi_{1\, {\rm min}}=\phi_1(n-1,n)
\ee
where one has to use Eqs. (\ref{solx1}) and (\ref{soly1}) to express $x$ and $y$, and
\be{max}
\phi_{1\, {\rm max}}=\phi_1(1,n).
\ee
The solution of the equation
\begin{equation}
\sum_{i=1}^{n} x_i^2 \ln x_i = L
\label{eqbase}
\end{equation}
exists on the constraint surface {\cal S} only if
\begin{equation}
\phi_{1\, {\rm min}} < L < \phi_{1\, {\rm max}}.
\label{maincondition}
\end{equation}
This constraint will be then used in the next section as a criterium for building the extensions of the SM containing a $750 \;\rm GeV$ boson.
Let us note that this last result is also valid when multiplicities $w_i\neq 1$ are considered. For such a case equations (\ref{system}-\ref{phi1}) are the same as $w_i$ simplifies. Depending on the particle content, however, $k$ may not take all the integers values in the interval $\paq{1,n-1}$ and the resulting (\ref{minim},\ref{max}) should be modified accordingly.\\
In paper \cite{we1} by direct calculation it was shown  that it was impossible to 
to satisfy the conditions (\ref{quad}), (\ref{log}), (\ref{en-finite1}) by adding some hypothetical bosons.
The extension of the Standard Model had  to also include  some new fermions.
In the next section we apply the technique sketched in this section to various hypothesis concerning the particle responsible for 
the 750 \rm{GeV} puzzle.

%%%%%%%%%%%%%%%%%%%%%%%
\section{Beyond the Standard Model: 750 {\rm GeV} puzzle and the constraint equations}

The Standard Model particles which give a significant contribution to our equalities have  the following particle masses: the top
quark mass $m_t =
173\  \rm{GeV}$, the Higgs mass $m_H = 125\  \rm{GeV}$, the  Z boson mass $m_Z = 91\  \rm{GeV}$ and the W boson mass 
$m_W = 80\  \rm{GeV}$. If we normalize $x_i$ w.r.t. the top mass we then have $x_t=1$, $x_H= 0.52$, $x_Z=0.28$, $x_W=0.21$. 

Let us suppose that  a scalar boson exists with  mass $m_B=750\  \rm{GeV}$ and $x_B= 18.8$. In this case the bosons dominate the constraint equations and 
\be{bal1}\left\{
\begin{array}{l}
\pa{R^{(1)}}^2=x_B^2+x_H^2+3x_Z^2+6x_W^2-12x_t^2\\
\\
h^{(1)}=x_B+x_H+3x_Z+6x_W-12x_t\\
\\
L^{(1)}=x_B^2\ln x_B+x_H^2\ln x_H+3x_Z^2\ln x_Z+6x_W^2\ln x_W-12x_t^2\ln x_t
\end{array}\right.
\ee
with $R^{(1)}=18.5$, $h^{(1)}=9.43$ and $L^{(1)}=1035$. Hence we wish to add fermions which provide the necessary cancellations. However $h^2 < R^2$ and the condition (\ref{posit}) is not satisfied hence it is impossible to find such a set of fermions without also adding some bosons.  One can then follow two strategies. One can first add a fermion $x_f$ which is massive enough to invert the balance, satisfy the conditions (\ref{posit}) and (\ref{maincondition}) and then search for a set of bosons satisfying our constraint equations which is the approach previously followed (\cite{we1}). Conversely one can add a massive boson $x_b$ with $n_b$ degrees of freedom such that $\pa{h+n_b x_b}^2>R^2+n_b x_b^2$ and then search for a set of fermions cancelling vacuum energy divergencies and finite part. In the latter case many massless boson degrees of freedom must be added to in order to compensate for the fermion ones,  hence we then prefer to follow the former approach, whenever possible. For both cases we assume that the massive particle which must be added is heavier than $x_B$ and is thus still unobserved.
%%%%%%%%%%%%%%%%%%%%%%%
\begin{figure}[t!]
\centering
\epsfig{file=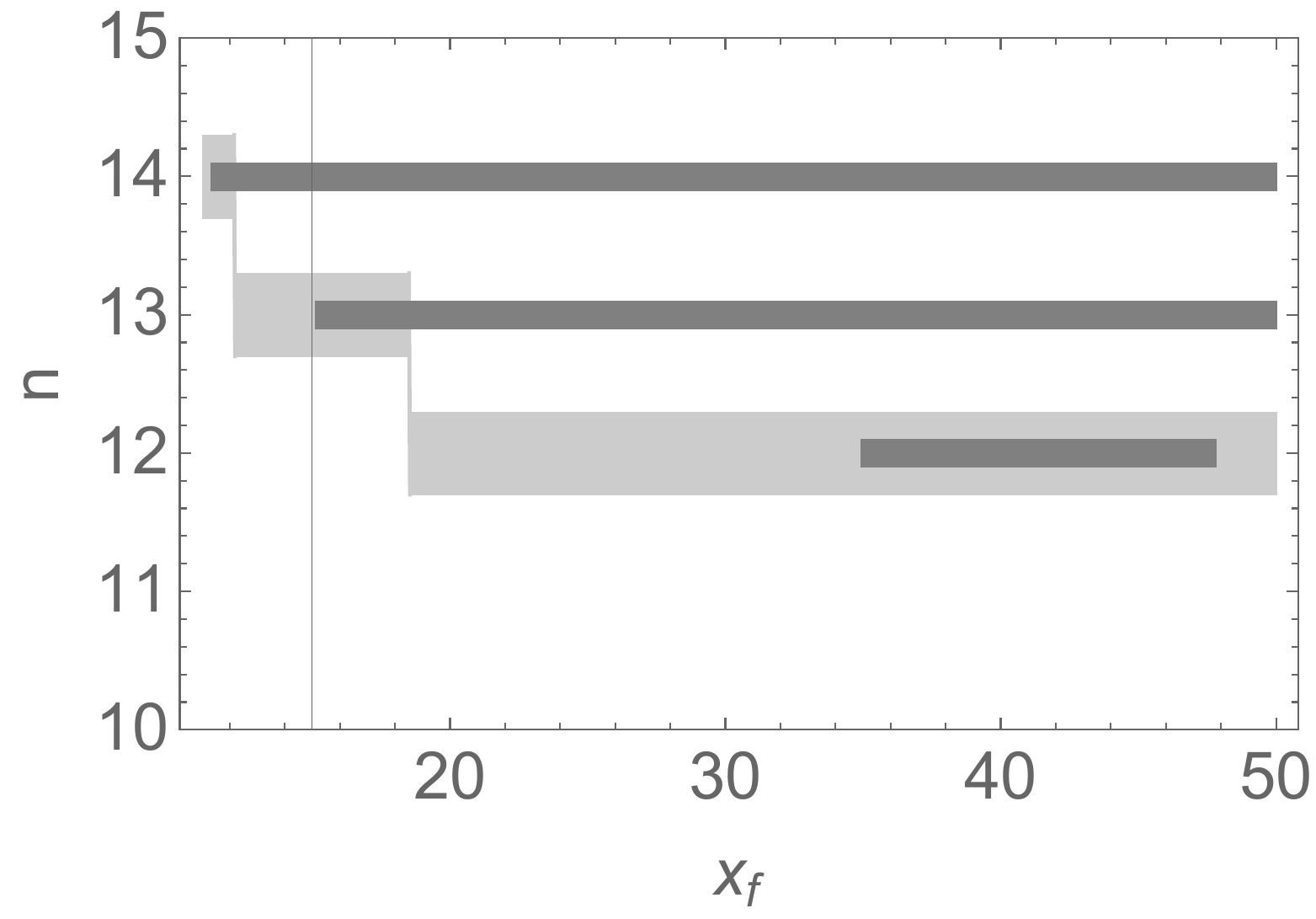, width=4.5 cm}
\epsfig{file=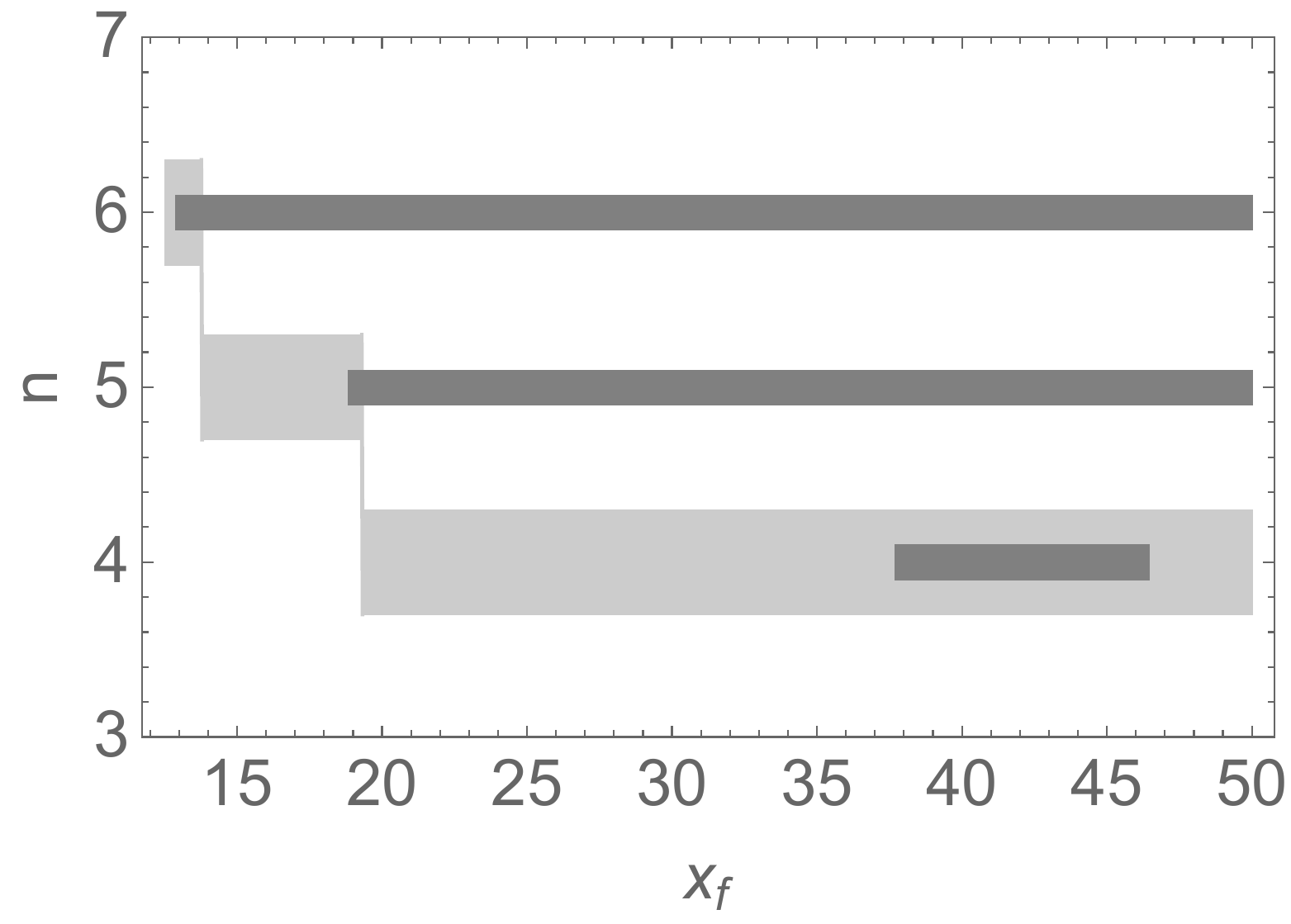, width=4.5 cm}
\epsfig{file=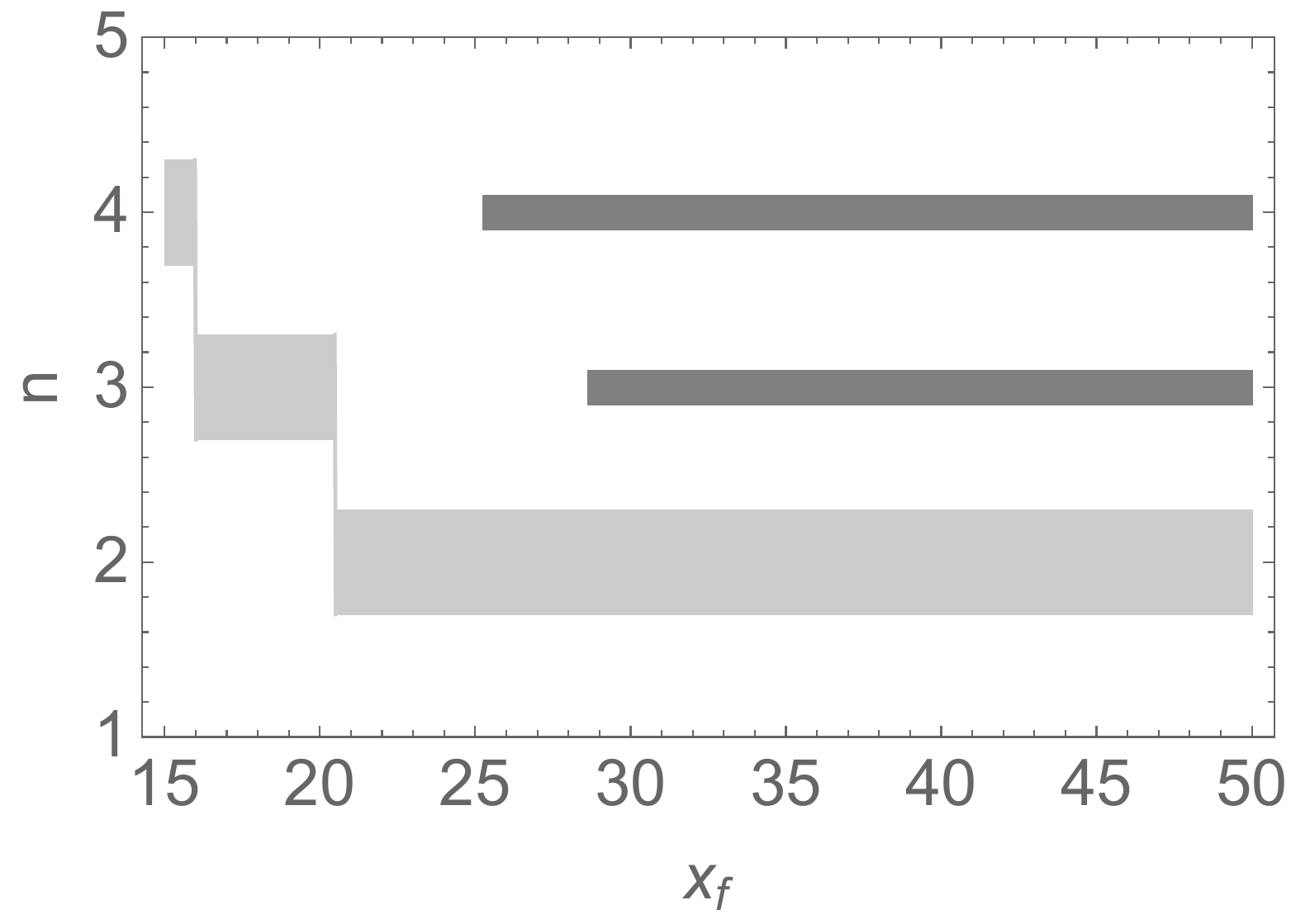, width=4.5 cm}
\caption{The above figures plot $n_0(x_f)$ (large, light grey band) and the constraint (\ref{maincondition}) on varying $x_f$ and for different choices of $n$ (small, dark grey band) when a $750\;GeV$ scalar and a fermion with mass $x_f$ are added to the SM. In the figure on the l.h.s. the case for a quark is plotted, in the figure on the centre the case with a colourless Dirac fermion is shown and in the figure on the r.h.s. the case for a Majorana fermion is shown. \it\label{fig1}}
\end{figure}
%%%%%%%%%%%%%%%%%%%%%%%%
\subsection{An extra coloured quark}
Let us first consider the case of a coloured quark, which has 12 degrees of freedom, with mass $x_f$. One can then plot  $n_0$ and the constraint (\ref{maincondition}) as functions of $x_f$ (see Fig. (\ref{fig1}) on the left). Thus the intervals for the quark mass which satisfy the constraint (\ref{maincondition}) are $x_f \in [34.9,47.8]$  ($m_f \in [1022,1196]$) with $12$ d.o.f., $x_f \in [15.1, 30.8]$ and $x_f\in [11.6, 26.9]$ with 13 and 14 d.o.f. respectively. We shall not consider a larger number of degrees of freedom even if it is possible, in principle, to satisfy  (\ref{maincondition}) with $n>14$ and $x_f>x_B$.\\
The first case given above -- adding12 boson degrees of freedom $x_i$ -- is relevant because $n=n_0$ and from (\ref{posmin}) we know , independently of the multiplicities $w_i$ of the boson particle content, for $x_f \in [34.9,47.8]$, that the solutions to the constraints must have $x_i>x_B$. Conversely if $n>n_0$ the geometrical minimum $x_m$ is less than $x_B$ and we expect solutions with a few masses below $x_B$.

\begin{figure}[t!]
\centering
\epsfig{file=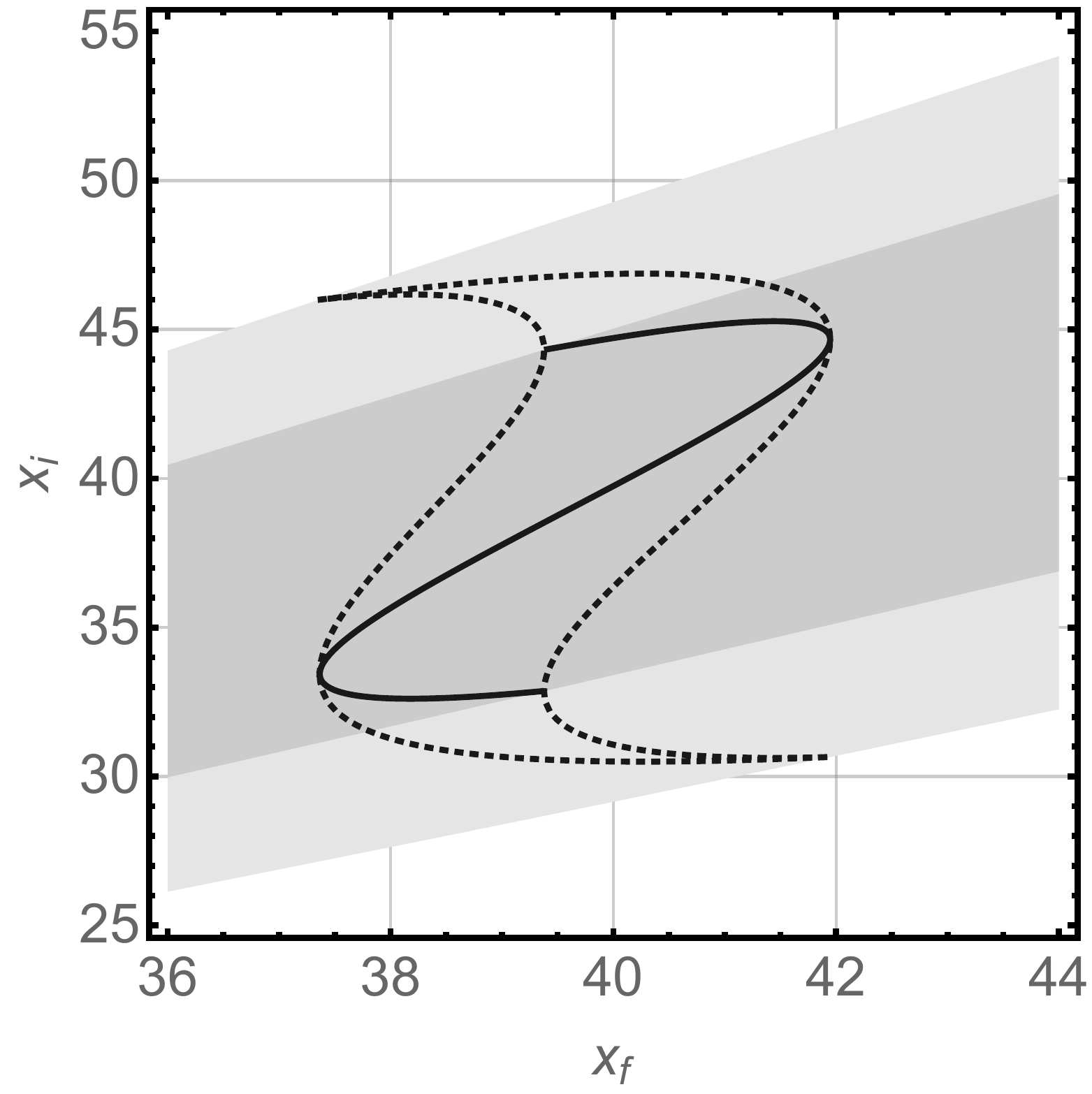, width=5.4 cm}
\epsfig{file=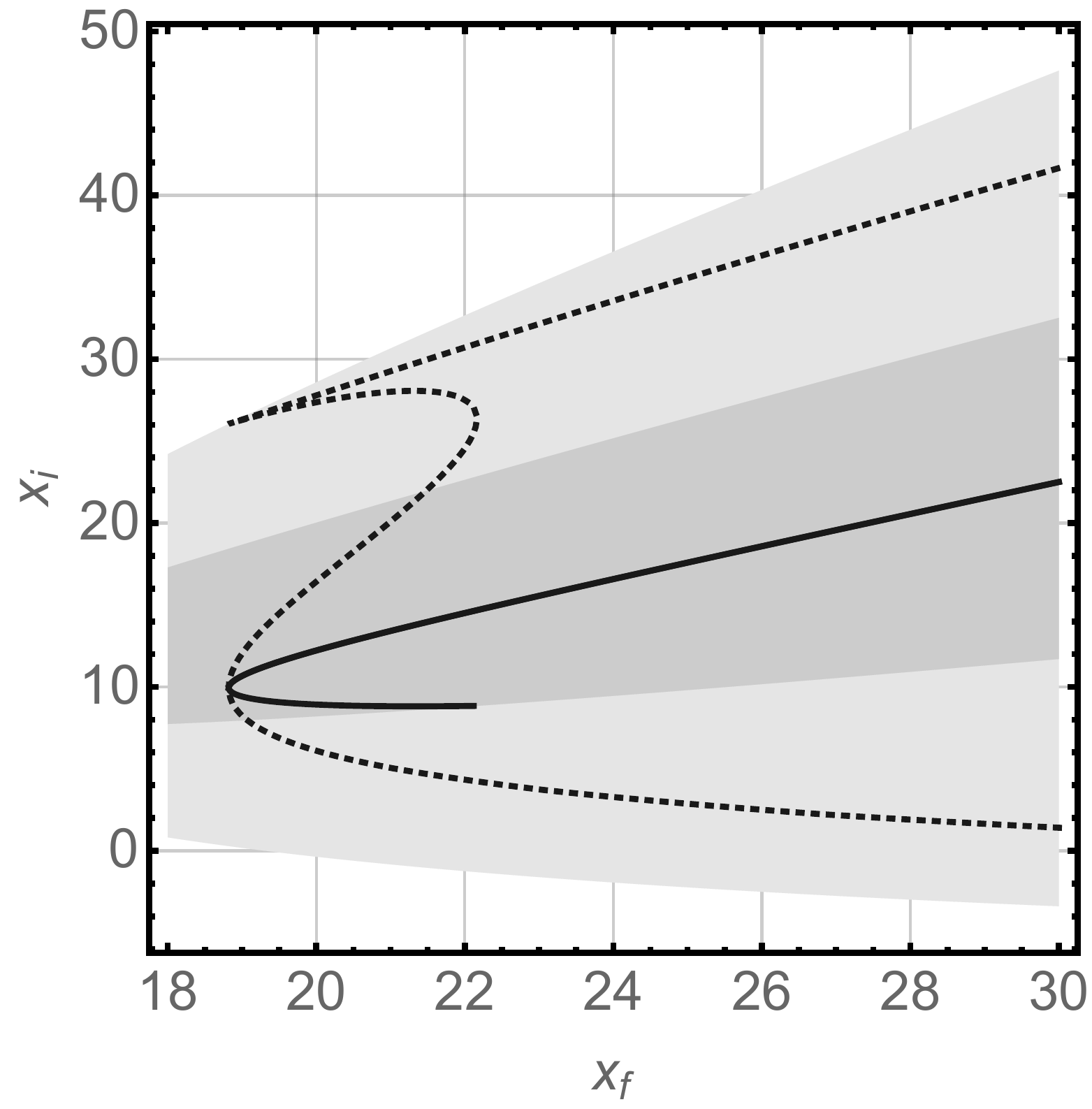, width=5.4 cm}
\caption{The figure on the l.h.s. is the one-parameter set of solutions of the case 2a); the solid line plots the mass $x_1$ and the dotted line represents the masses $x_2$ and $x_3$. The figure on the r.h.s. is  the one-parameter set of solutions of the case 2b); the solid line plots the mass of the vector $x_1$ and the dotted line represents the masses $x_2$ and $x_3$ of the scalars. The shaded areas represent the region between $x_m$ and $x_M$ defined by (\ref{minmax}).\it\label{fig2}}
\end{figure}

We first study a few cases with the minimal particle content. Such cases can be investigated numerically and the entire set of solutions to the constraints can be easily plotted. 
\begin{itemize}
\item[1a)] Consider the addition of 4 vector bosons with two independent masses $x_1$ and $x_2$. For such a case we satisfy the three constraints for $x_f=32.9, x_1=32.9$ and $x_2=44.3$ and these 3 particles are heavier than $750\ \rm{GeV}$.
\item[2a)] Consider the addition of 4 vector bosons with 3 independent masses $x_1$, $x_2$ and $x_3$ where  $x_1$ is the common mass of 2 vector bosons. One then finds a one-parameter set of solutions shown on the left-hand side of Fig. (\ref{fig2}) where :
the dotted line represents the coincident masses of two vector bosons, while the solid line represents the masses of the remaining bosons. All the boson and fermion masses are heavier than $750\ \rm{GeV}$.
\end{itemize}
More general cases with at least 12 bosonic degrees of freedom can also be discussed as the minimum masses in the multiplet always appear in patterns made of 3 distinct $\bar x_i$ (as discussed in the previous section). More features of such cases can be extracted numerically by using a Monte Carlo (MC) inspired technique which  samples the hyperspace  spanned by the masses in the boson sector $x_i$. 
If we consider a set of bosons containing 3 new massive vectors and 4 scalars and fix $x_f=45$ ($m_f=1160\ \rm{GeV}$),  we can then calculate the lightest mass of the scalar sector $x_s=24.4$ , of the lightest vector $x_v=41.3$ and with the MC we find their respective mass interval $m_s\in [854,867]\ \rm{GeV}$, $m_v\in [1112,1167]\ \rm{GeV}$ . \\
If two of these three vector bosons have the same mass (like the $W^{\pm}$ bosons of the Standard Model), then the lightest scalar particle has a mass in the interval $m_s\in [854,871]\ \rm{GeV}$, while the lightest vector particle has a mass in the interval $m_v \in [1113,1191]\ \rm{GeV}$. \\

%%%%%%%%%%%%%%%%%%%%%%%%%%
%%%%%%%%%%%%%%%%%%%%%%%%%%
\subsection{An extra Dirac fermion}

Let us now consider the case of an additional Dirac fermion with 4 degrees of freedom. On plotting $n_0$ and the constraint (\ref{maincondition}) as functions of $x_f$ (see Fig. \ref{fig1} in the middle) one observes that at least 4 boson degrees of freedom are needed (with $x_f\in[37.7,46.4]$), 5 if $x_f\in[18.8,19.3]$.  Starting with the cases with a minimal particle content we find the following.\\
\begin{itemize}
\item[1b)] On adding one vector boson ($x_v$) and one scalar ($x_s$) : for this case we satisfy the constraints with $x_f=37.7, x_v=29.9$ and $x_s=51.4$ with the three particles heavier than $750\rm{GeV}$.\\
\item[2b)] On adding a vector boson and 2 scalars ($x_1$, $x_2$ and $x_3$ respectively), we find a one-parameter set of solutions illustrated in the Fig. (\ref{fig2}). For such a case light particles are always present in the spectrum.\\
\item[3b)] On adding a Dirac fermion with mass $x_f=45$ and four scalar particles with different masses, one has that the lightest mass in the scalar sector is $m_s =854 \;\rm{GeV}$ and the boson multiplet is made of four particles particles heavier than $750\rm{GeV}$. 
\end{itemize}

\subsection{An extra Majorana fermion}
On adding a Majorana fermion with 2 degrees of freedom, we find that it is necessary to have at least two additional boson d.o.f. to satisfy (\ref{posit}). The cases which we considered are the following :\\
\begin{itemize}
\item[1c)] On adding two scalar fields $x_1$ and $x_2$ we  find the solution $x_f=43.9, x_1=24.4, x_2=54.1$ with the three particles heavier than $750\rm{GeV}$.\\
\item[2c)] On adding a vector boson and a scalar ($x_v$ and $x_s$ respectively), we find $x_f=25.2$, $x_s=29.8$, $x_v=3.7$, and a light boson in the spectrum.\\
\end{itemize}
%%%%%%%%%%%%%%%%%%%%%%%%%%
\begin{figure}[t!]
\centering
\epsfig{file=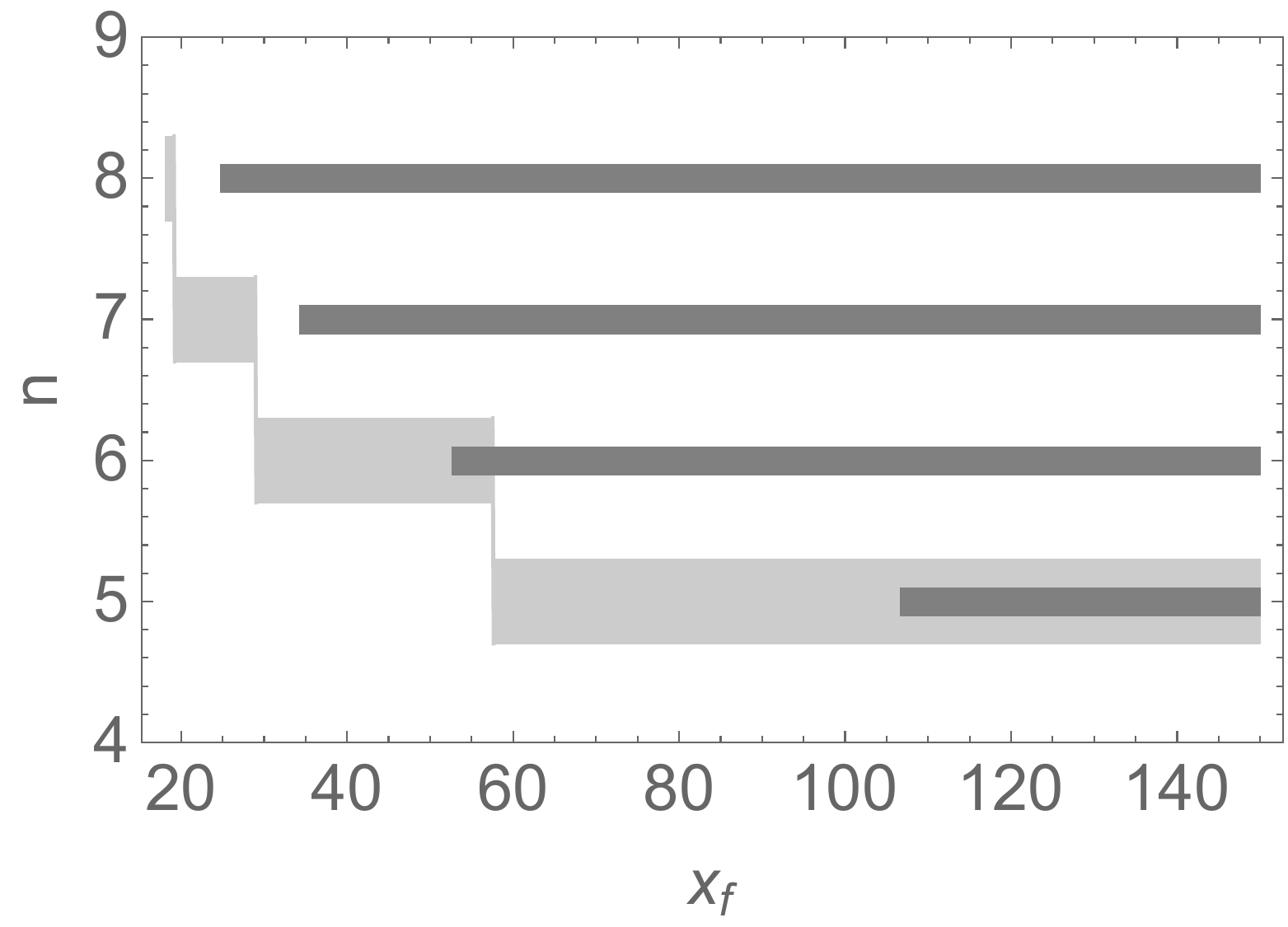, width=5.5 cm}
\epsfig{file=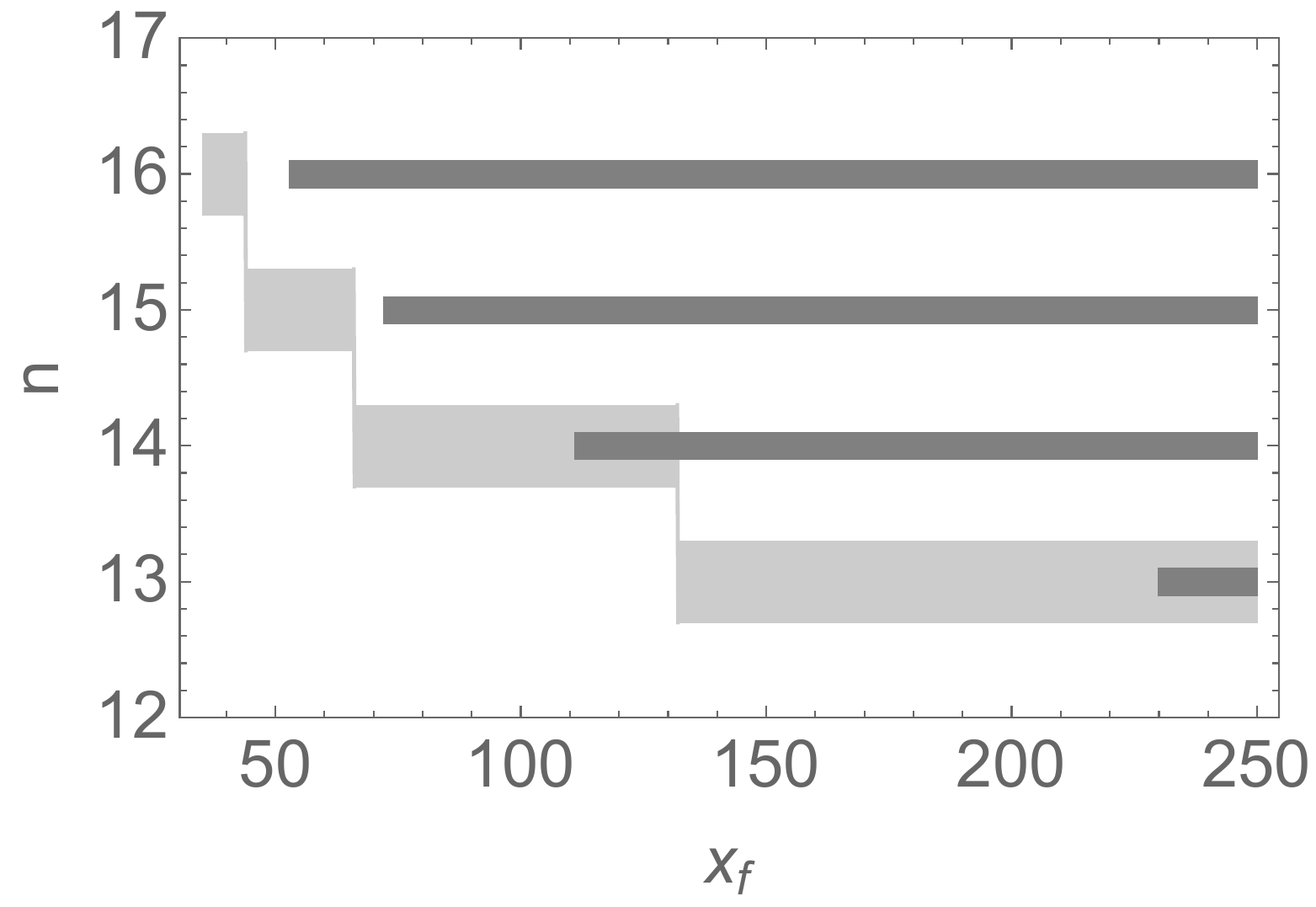, width=5.5 cm}
\caption{The above figures plot $n_0(x_f)$ (large, light grey band) and the constraint (\ref{maincondition}) on varying $x_f$ and for different choices of $n$ (small, dark grey band) when a $375\;GeV$ fermion and a fermion with mass $x_f$ are added to the SM. In the figure on the l.h.s. the case with a two Dirac fermions is plotted, i and in the figure on the r.h.s. the case with two quark fermions is shown. \it\label{fig1b}}
\end{figure}
\subsection{Composite particle}
\begin{figure}[t!]
\centering
\epsfig{file=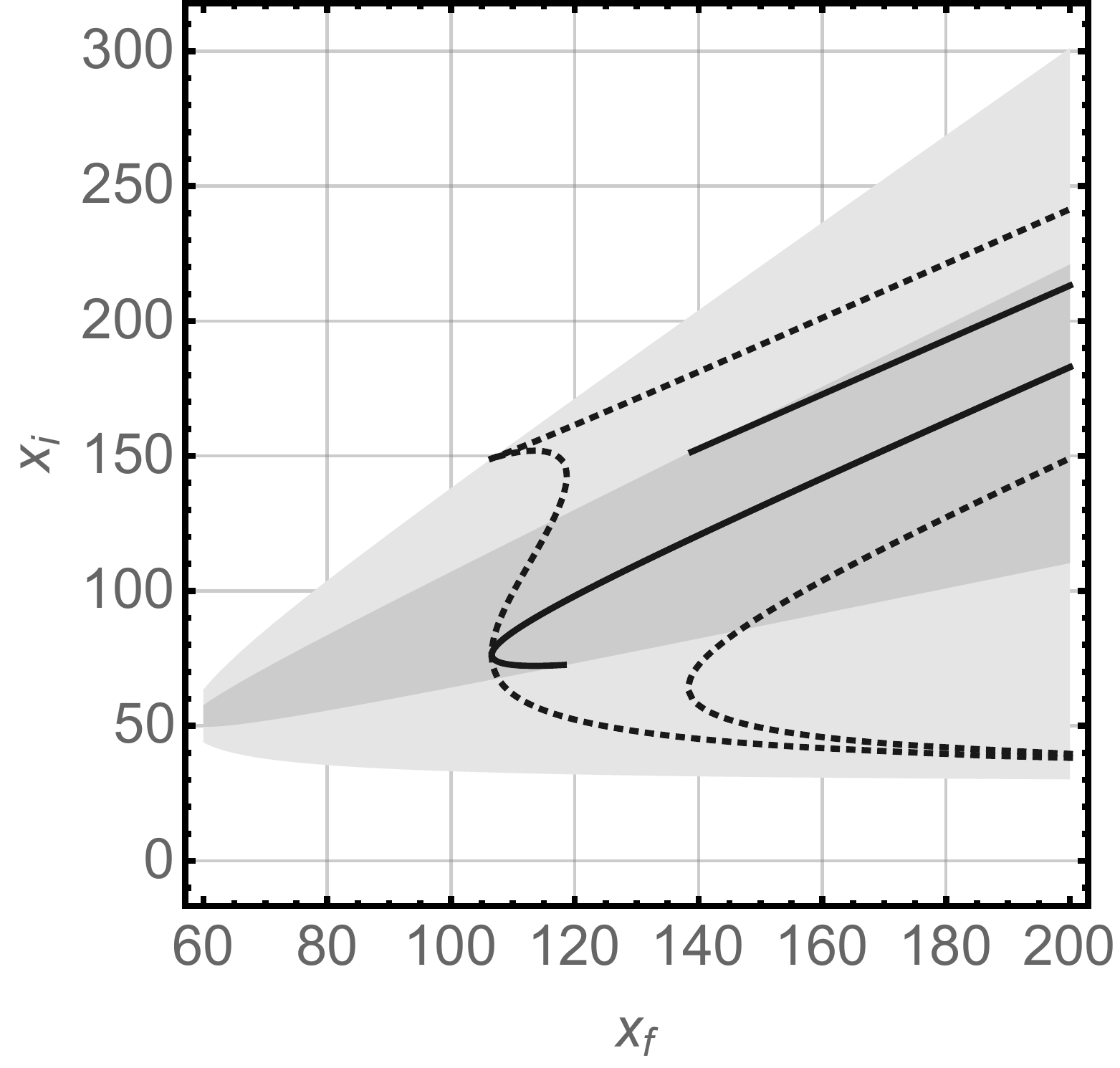, width=5.4 cm}
\epsfig{file=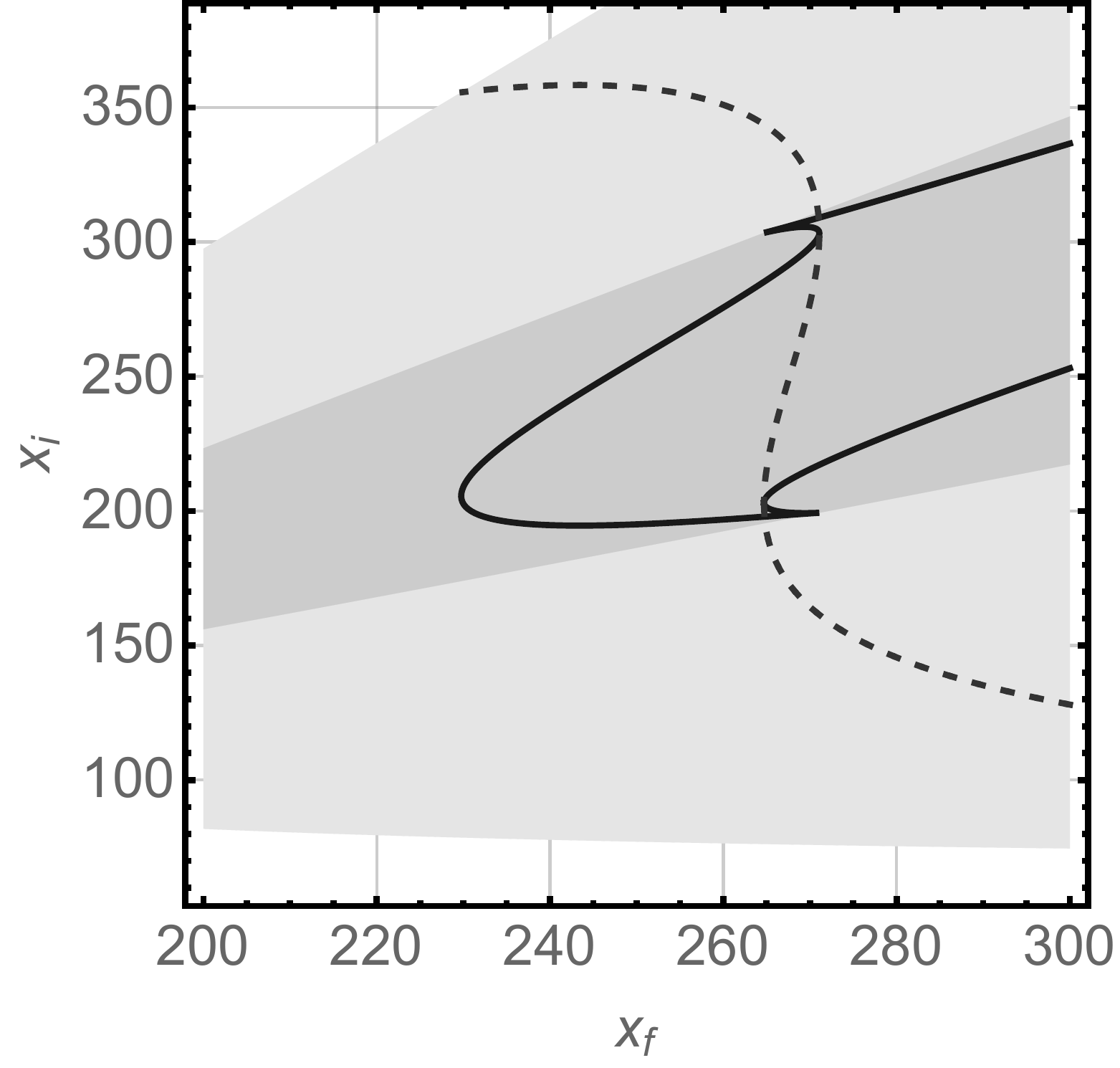, width=5.4 cm}
\caption{The figure on the l.h.s. is  the one-parameter set of solutions of the case 2c); the solid line plots the mass $x_1$ and the dotted line represents the masses $x_2$ and $x_3$. The figure on the r.h.s. is the one-parameter set of solutions of  the case 3c); the solid line plots the mass of the vectors $x_1$ and $x_2$ and the dotted line represents the mass of the scalar $x_3$. The shaded areas represent the region between $x_m$ and $x_M$ defined by (\ref{minmax}).\it\label{fig2b}}
\end{figure}

Instead of the $750\ \rm{GeV}$ scalar particle one can suppose that   the  
750 \rm{GeV} diphoton excess at the LHC is associated with the existence of a composite boson made of a Dirac fermion-antifermion pair with  a total mass $750\ \rm{GeV}$ and negligible binding energy. For such a case the vacuum energy cancellation constraints are
\be{bal2}\left\{
\begin{array}{l}
\pa{R^{(c)}}^2=-w_f x_f^2+x_H^2+3x_Z^2+6x_W^2-12x_t^2\\
\\
h^{(c)}=-w_f x_f+x_H+3x_Z+6x_W-12x_t\\
\\
L^{(c)}=-w_fx_f^2\ln x_f+x_H^2\ln x_H+3x_Z^2\ln x_Z+6x_W^2\ln x_W-12x_t^2\ln x_t
\end{array}\right.
\ee
with $R^{(c)}=16.6$, $h^{(c)}=65.7$ and $L^{(c)}=410.8$ for the case of a coloured Dirac fermion and $R^{(c)}=10$, $h^{(c)}=28.2$ and $L^{(c)}=137.5$ for the case of a colourless Dirac fermion. In both cases the condition (\ref{maincondition}) is not satisfied for $n=n_0$ and one must add one more fermion in order to keep the masses of the boson d.o.f. higher than $750\; {\rm GeV}$.  
Let us consider the case of colourless fermions. In this case it is necessary to add one more Dirac fermion. For such a  case one also needs at least 5 massive boson d.o.f. (see Fig. \ref{fig1b})). Starting with just a few particles we have
\begin{itemize}
\item[1c)] Addition of two vector bosons ($x_1$ and $x_2$). In this case we satisfy the constraints with $x_f=73.1, x_1=26.6$ and $x_2=81.2$.\\
\item[2c)] For a vector boson and 2 scalars ($x_1$, $x_2$ and $x_3$ respectively), we find a one-parameter set of solutions illustrated in the Fig. (\ref{fig2b}). For such a case the particles needed are always heavier than $750\;{\rm GeV}$.
\end{itemize}
The cases with more particles involved can be studied separately. 
Let us take a fermion of  mass $x_f=110$. Then, one has to add at least 5 boson degrees of freedom. On
considering 5 scalar fields, we find that the lightest one has a mass in the interval 
$m_s \in [1359,1469]\ \rm{GeV}$. \\ 
The composite boson can also be made  of a coloured quark-antiquark pair. 
This case is similar to the colourless case but now at least 13 d.o.f. are involved.  
\begin{itemize}
\item[3c)]If we consider the case of 4 vector bosons with 2 different masses $x_1$ and $x_2$ plus a scalar $x_3$, we find a one-parameter set of solutions illustrated in  Fig. (\ref{fig2b}). 
\end{itemize}
Let us consider a quark with mass $x_f=120$ and 14 scalars. For this case the lightest scalar in the multiplet has a mass  $m_s=1268\ \rm{GeV}$. The same lightest scalar appears in more complicated multiplets with, at least, 2 scalars. For example let us consider a multiplet formed by three vector fields (two of them with the same mass) and 5 scalars. In this case the lowest  scalar mass lies in the interval $m_s \in [1268,1777]\ \rm{GeV}$ while the lowest vector mass belongs to the interval $m_v \in [1483,1815]\ \rm{GeV}$, where the first value in the last two intervals is calculated analytically and the second is estimated by MC techniques. 
%%%%%%%%%%%%%%%%%%%%%%%%%%%%%%%%%%%%%%%%%%%%%%%%%%%%
\subsection{Spin two particle}
Finally we consider a spin two particle - a massive graviton, which has 5 degrees of freedom (see e.g. \cite{massive}). The  equations (\ref{constraint1}-\ref{constraint3}) take the following form:
\be{bal3}\left\{
\begin{array}{l}
\pa{R^{(2)}}^2=5x_B^2+x_H^2+3x_Z^2+6x_W^2-12x_t^2\\
\\
h^{(2)}=5x_B+x_H+3x_Z+6x_W-12x_t\\
\\
L^{(2)}=5x_B^2\ln x_B+x_H^2\ln x_H+3x_Z^2\ln x_Z+6x_W^2\ln x_W-12x_t^2\ln x_t
\end{array}\right.
\ee
with $R^{(2)}=41.9$, $h^{(2)}=84.6$ and $L^{(2)}=5180$.
To balance its big contribution to the vacuum energy, one has add a fermion with a mass greater than that of the massive graviton. For such a minimal case we could not find any interesting solution (with masses bigger than $750\;{\rm GeV}$)  on only adding one or two massive fermions, which does appear satisfactory from the observations point of view. \\

%%%%%%%%%%%%%%%%%%%%%%%%%%%%%%%%%%
\section{Conclusions}
%%%%%%%%%%%%%%%%%%%%%%%%%%%%%%%%%%
 Let us begin by briefly summarizing our results.  
 We first considered the case for which the $750\;{\rm GeV}$ particle is a scalar and found, in order to satisfy the constraints, the following minimal particle contents:
\begin{itemize}
\item[a)] an extra coloured quark with mass in the interval $[1022,1196]\;{\rm GeV}$ and correspondingly 12 bosonic d.o.f.;

\item[b)] an extra Dirac fermion with mass in the interval  $[1062,1178]\;{\rm GeV}$ and correspondingly 4 bosonic d.o.f.;

\item[c)] an extra Majorana fermion with a mass of $1146\;{\rm GeV}$ and correspondingly 2 scalar d.o.f..
\end{itemize}

Going beyond such minimal cases one generally finds unsatisfactory (light) masses for either the fermion or the bosons.\\
We then considered the case for which the $750\;{\rm GeV}$ particle is a composite object consisting of a lightly bound fermion-antifermion pair, each fermion having a mass of $375\;{\rm GeV}$. In order to satisfy the constraints we found the following minimal particle contents:

\begin{itemize}
\item[d)] for the $375\;{\rm GeV}$ Dirac fermion, one extra colourless fermions and at least 5 bosonic d.o.f.;

\item[e)] for the $375\;{\rm GeV}$ coloured fermion, one extra coloured quark and at least 13 bosonic d.o.f..
\end{itemize}
Lastly we considered a spin 2 $750\;{\rm GeV}$ particle. We could not compensate such a large contribution on only adding one or two extra massive fermions plus any massive boson (heavier than $750\;{\rm GeV}$) multiplet.

We have thus seen that on adding to the Standard Model different hypothetical particles associated with the 750 \rm{GeV} diphoton excess and requiring the cancellation of its vacuum energy, one can find  solutions in which  various new fields appear, both bosons and fermions with  masses compatible with present data. 

Let  us further  note that to treat 
the finite contribution of quantum fields to the  vacuum energy it is reasonable to combine 
the cancellation mechanism proposed by W. Pauli \cite{Pauli} which was applied in \cite{we,we1} and in the present 
paper together with the standard renormalization procedure for the ultraviolet divergences in the 
traditional quantum field theory of the scattering matrix (see e.g. \cite{Bog-Shir,Nambu}).   
Indeed, the cancellation of the contributions of boson and fermion fields to the vacuum energy 
is useful only to treat the one-loop divergences coming from propagators of free fields. 
As is well-known these divergences, which in standard quantum field theory,  in the 
absence of gravity, are eliminated by the normal ordering procedure, create problems, when 
gravity is included.  All the other ultraviolet divergences need not be cancelled, but rather
renormalized. It is worth remembering that even in the majority of models possessing an exact supersymmetry 
not all the ultraviolet divergences are cancelled. For example, in the Wess-Zumino model \cite{WZ} 
the loop-contributions to the vertices are ultraviolet finite, while those to the 
propagators are divergent and must be renormalized. 

Thus, the standard procedure of the elimination of  ultraviolet divergences, results in the 
renormalization of masses, coupling constants and  kinetic terms of the fields under consideration. 
The last (kinetic terms) can be fixed in such a way that they have standard form, i.e. the corresponding 
renormalization coefficients are equal to one \cite{Bog-Shir}. The renormalization of the masses reduces 
 to the appearance of the physically measurable masses - these are the ones which we used 
for the sum rules in \cite{we,we1} and in the present paper. \\

Let us conclude by noting that we have assumed some
suitable breaking of supersymmetry which breaks equality of boson and
fermion masses (existing before breaking even in the presence of interactions), but still keeps the relations (\ref{quad}), (\ref{log}),(\ref{en-finite1}). 
It could be that
this is possible in the context of supergravity only. This is because
we know that even in renormalizable theories, only "measurable"
quantities participating in some interactions can be renormalized,
quantitites which we cannot measure using some interaction vertices
need not be finite. Since gravity is needed to measure the total
vacuum energy density, not its difference between different quantum
states, gravity has to be included, and then supersymmetry transforms
to supergravity. However this is a scope for further work.

%%%%%%%%%%%%%%%
\section*{Acknowledgements}
%%%%%%%%%%%%%%%
The work of A.K. and A.S.  was partially supported by the RFBR grant  14-02-00894.

%%%%%%%%%%%%%%%

%%%%%%%%%%%%%%%

\end{document}